\documentclass[manuscript,screen]{acmart}
\usepackage{amsthm}
\usepackage{booktabs}
\usepackage{algpseudocode}
\usepackage{algorithmicx}
\usepackage{algorithm}
\usepackage{subfigure}
\usepackage{color}
\usepackage{multirow}
\newcommand\sbullet[1][.5]{\mathbin{\vcenter{\hbox{\scalebox{#1}{$\bullet$}}}}}
\usepackage{wasysym}

\AtBeginDocument{%
  \providecommand\BibTeX{{%
    \normalfont B\kern-0.5em{\scshape i\kern-0.25em b}\kern-0.8em\TeX}}}

\setcopyright{acmcopyright}
\copyrightyear{2018}
\acmYear{2018}
\acmDOI{XXXXXXX.XXXXXXX}

\begin{document}

\title{COMET: Convolutional Dimension Interaction for Collaborative Filtering}

\author{Zhuoyi Lin}
\authornote{This work was done when he was a student at School of Computer Science and Engineering, Nanyang Technological University, Singapore.}
\email{Lin_Zhuoyi@i2r.a-star.edu.sg}
\affiliation{%
  \institution{School of Computer Science and Engineering, Nanyang Technological University, and Institute for Infocomm Research (I$^{2}$R), Agency for Science, Technology and Research (A*STAR)}
  \country{Singapore}
}

\author{Lei Feng}
\authornote{Corresponding author.}
\email{feng0093@e.ntu.edu.sg}
\affiliation{%
  \institution{School of Computer Science and Engineering, Nanyang Technological University}
  \country{Singapore}
}

\author{Xingzhi Guo}
\email{xingzguo@cs.stonybrook.edu}
\affiliation{%
  \institution{Department of computer science, Stony Brook University}
  \country{USA}
}

\author{Yu Zhang}
\email{yu.zhang@icr.ac.uk}
\affiliation{%
  \institution{The Institute of Cancer Research}
  \country{UK}
}

\author{Rui Yin}
\email{ruiyin@ufl.edu}
\affiliation{%
  \institution{Department of Health Outcomes and Biomedical Informatics, University of Florida}
  \country{USA}
}

\author{Chee Keong Kwoh}
\email{asckkwoh@ntu.edu.sg}
\affiliation{%
  \institution{School of Computer Science and Engineering, Nanyang Technological University}
  \country{Singapore}
}

\author{Chi Xu}
\email{cxu@simtech.a-star.edu.sg}
\affiliation{%
  \institution{Singapore Institute of Manufacturing Technology (SIMTech), Agency for Science, Technology and Research (A*STAR); and School of Computer Science and Engineering, Nanyang Technological University}
  \country{Singapore}
}

\renewcommand{\shortauthors}{Zhuoyi Lin, et al.}

\begin{abstract}
Representation learning-based recommendation models play a dominant role among recommendation techniques. However, most of the existing methods assume both historical interactions and embedding dimensions are independent of each other, and thus regrettably ignore the high-order interaction information among historical interactions and embedding dimensions.
In this paper, we propose a novel representation learning-based model called COMET (\underline{CO}nvolutional di\underline{M}\underline{E}nsion in\underline{T}eraction), 
which simultaneously models the high-order interaction patterns among historical interactions and embedding dimensions.
To be specific, COMET stacks the embeddings of historical interactions horizontally at first, which results in two "embedding maps". 
In this way, internal interactions and dimensional interactions can be exploited by convolutional neural networks (CNN) with kernels of different sizes simultaneously. 
A fully-connected multi-layer perceptron (MLP) is then applied to obtain two interaction vectors.
Lastly, the representations of users and items are enriched by the learnt interaction vectors, which can further be used to produce the final prediction. 
Extensive experiments and ablation studies on various public implicit feedback datasets clearly demonstrate the effectiveness and rationality of our proposed method.
\end{abstract}

\ccsdesc[500]{Computer systems organization~Embedded systems}
\ccsdesc[300]{Computer systems organization~Redundancy}
\ccsdesc{Computer systems organization~Robotics}
\ccsdesc[100]{Networks~Network reliability}

\keywords{recommender systems, implicit feedback, interaction modeling, representation learning}

\maketitle

\section{Introduction}
In the era of big data, analyzing customers' demands and behaviors is necessary to exploit potential insights and build intelligent systems, which can be achieved by modern recommender systems.
Suggestions for videos on YouTube or products on Amazon \cite{covington2016deep,linden2003amazon,wang2018serendipitous}, are real-world examples of intelligent systems to help users navigate a growing ocean of choices.
To this end, collaborative filtering (CF) methods are proposed to estimate
users' preferences from her historical behaviors, which are widely adopted due to their impressive recommendation performance.
For example, representation learning-based models which aim to learn effective users' representations and items' representations using deep learning techniques have played a dominant role in recent years \cite{NCF,weimer2008cofi,candes2009exact,liu2019matrix,lin2021glimg}. The matching score between the target user and the target item can be predicted by leveraging their representations.

In order to effectively capture the latent relationships between users and items, the recommender system research community and industry have paid great attention and efforts to model the interaction information between contextual features. 
A typical solution is to model the domain-specific cross features manually \cite{cheng2016wide,wang2018tem}. For example, cross-product transformation is used over sparse features to encode feature interactions in \cite{cheng2016wide}. Although these methods are able to discover the relationships between feature pairs and target labels in an explicit manner, tedious efforts are required to construct cross features and such feature interaction cannot generalize to unseen cross-features. 
Alternatively, the recommendation models could learn feature interactions automatically. 
In particular, Factorization Machine (FM) and its extensions \cite{rendle2010factorization,he2017neural} map contextual features to low-dimensional embeddings. This enables the interaction between contextual features can be estimated by the inner product of their embeddings.
For example, the interaction between users' gender and items' categories are recognized as second-order interaction.
However, most of the FM-based methods only model pairwise (second-order) feature interactions due to the computational inefficiency of explicitly enumerating high-order feature interactions (i.e., the interaction among more than three features).

Another attempt to model interaction information is to capture dimension-level interaction information. The motivation is to treat the latent factors encoded in the embeddings as user/item features \cite{qu2016product,he2018outer,xin2019cfm,koren2009matrix}.
For example, a user's representation may encode her gender, her spending power, and her preferred color. Similarly, an item can be characterized as male-oriented versus female-oriented, its price and its color. 
Recent works use outer-product to model the interaction between input pairs, and then multi-layer perceptron (MLP) \cite{qu2016product} or convolutional neural network (CNN) \cite{he2018outer,xin2019cfm} is used to estimate matching scores.
By using an outer-product operation on the target user's representation and the target item's representation, the generated interaction map explicitly encodes all the pairwise dimensional interactions. 
In this way, CNN is able to capture the dimensional interactions in a more explicit way compared to methods which directly employ MLP on embedding concatenation such as Deep Crossing \cite{shan2016deep,he2018outer,xin2019cfm}. 
However, the input of such an interaction modeling process is only pairwise, in other words, the outer-product can only encode the explicit interactions between only two dimensions, which may potentially ignore the rich information among latent embeddings.

Based on the above observations, we can see that both types of methods for modeling interaction information are deeply mired in the difficulty of explicitly modeling high-order interactions: most of the existing works either consider pairwise (second-order) feature interactions or pairwise dimension-level interactions.
Furthermore, most of the FM-based methods that aim to model the feature interaction information, are based on contextual features (e.g., item descriptions, rating, and user check-in data) which are not always available. 
In contrast, implicit feedback (e.g., click, browse, or purchase behaviors), is much easier to be collected \cite{rendle2009bpr,NCF,he2018outer}.
In this paper, we propose a novel approach COMET (\underline{CO}nvlutional di\underline{M}\underline{E}nsion in\underline{T}eraction) to simultaneously capture high-order interaction information among historical interactions and embedding dimensions from implicit feedback.
To be specific, we treat the interacted items and interacted users (i.e. items purchased by the target user and users who have consumed the target item) as  "contextual features" in this work. 
By stacking such historical interactions horizontally, two "embedding maps" can be obtained.
For each embedding map, we employ a single-layer CNN with kernels of variant sizes over it, which aims to capture high-order interaction signals among historical interactions and all embedding dimensions simultaneously.
A fully-connected MLP is then employed to achieve two interaction vectors.
By enriching the original representations of the target user and target item with such interaction vectors, our proposed method is able to obtain an impressive performance. 
In summary, the main contributions of this work are:
\begin{itemize}
\item {We propose a novel approach COMET to capture the interaction signals for recommendation from implicit feedback. COMET aims to exploit the high-order interaction information among historical interactions and embedding dimensions simultaneously.
}
\item {We propose to enrich the representations of the target user and the target item by the learnt interaction information. In this way, the target user's representation and the target item's representation are dependently learnt.} 
\item {We conduct extensive experiments on public implicit feedback data to evaluate the performance of our proposed method. Experimental results show that our proposed method is able to achieve impressive results. Moreover, ablation studies are conducted to analyze the advantages of COMET.} 

\end{itemize}
The rest of the paper is organized as follows: In section 2, related works are briefly reviewed. We then elaborate on our method in Sections 3 and Section 4. In Section 5, we empirically evaluate our proposed method on recommendation tasks. We conclude our work and discuss future directions in Section 6.

\section{Related Work}
Our work is built on the foundation of the latent factor models, and representation-based models and takes advantage of feature interaction modeling.

Latent factor models learn users’ and items’ latent embeddings in a shared latent space. These methods use low-rank approximation to fit the rating matrix. 
For example, In 2010, Karatzoglou et al. introduced a technique called tensor factorization (TensorF) \cite{karatzoglou2010multiverse} that allows for the incorporation of multiple features into a recommendation model. This is done by representing the data as a multi-dimensional tensor rather than a traditional 2D matrix. Based on TensorF, Symeonidis et al. \cite{symeonidis2010unified} developed a recommendation model called HOSVD that utilizes tensor factorization for user-tag-item triplet data. More recently, Yu et al. proposed a tensor factorization model called DCFA \cite{yu2018aesthetic} which uses aesthetic features, rather than traditional features, to make recommendations based on a user's preferences. They believed that a user's decision is often influenced by whether the product aligns with their personal aesthetics.
As special cases of tensor factorization, matrix factorization techniques (MF) \cite{koren2009matrix,rendle2009bpr} factorize the rating matrix into user-specific and item-specific matrices for rating prediction.
Another notable latent factor model is SVD++ \cite{svd}, which integrates the embedding of the target user with additional latent embeddings of interacted items.

Intuitively, the relationships between users and items are complex, thus representation-based recommendation models are proposed to learn the complex matching function that maps user-item pairs to matching scores.
For example, Generalized Matrix Factorization (GMF) \cite{NCF} is proposed to generalize MF in a non-linear manner. 
NeuMF \cite{NCF} and DeepCF\cite{deepcf} use MLP to learn effective matching functions from user/item representations or user/item rating data, respectively.
In recent years, many representation learning-based sequential recommendation methods are proposed due to the impressive ability of deep learning to learn the complex behaviors of users \cite{sun2019bert4rec,tang2018personalized,kang2018self,zhou2018deep,hidasi2015session}.
Specifically, sequential recommendation methods are used to model users' dynamic preferences and make personalized recommendations to users based on their sequential interactions with items.
For example, GRU4rec \cite{hidasi2015session} is proposed to process a sequence of items and make recommendations based on the hidden state learned by the gated recurrent unit (GRU).
In addition, DIN \cite{zhou2018deep} and SASRec \cite{kang2018self}  use attention mechanisms to weigh the importance of different items in a user's interaction history, which process a sequence of items and makes recommendations based on the learned attention weights and item representations.
Meanwhile, Caser \cite{tang2018personalized} is a  CNN-based method, which aims to model the skip behaviors of sequential patterns.
To utilize the pre-trained language model BERT for recommendation tasks, BERT4Rec \cite{sun2019bert4rec} first fine-tunes BERT on a large dataset of user-item interactions and then uses the resulting fine-tuned model to make recommendations based on a given sequence of items.
However, the motivation and experimental settings of sequential methods are different from ours. We thus omit the comparison with sequential recommendation models in this work. 
To be specific, COMET and other general recommendation methods \cite{rendle2009bpr, svd, fism,he2018outer, NCF,he2020lightgcn} aim to learn static and inherent user preferences from their historical data, which recommend items to users based on their overall preferences and interests. Note that these methods do not take into account the order in which the items were consumed or the time at which they were consumed \cite{sun2019research}. 
Sequential recommendation methods, on the other hand, take into account the sequence of items that have been consumed by a user. 
These methods learn the sequential information from sorted historical interactions, which model dynamic user preferences that change from time to time \cite{sun2019bert4rec,tang2018personalized,kang2018self}.
Therefore, we only compare baselines that aim to model static user preferences based on their unsorted historical interactions in the paper as we focus on modeling general user preferences.
Above all, we can observe that most of the existing latent factor approaches and representation-based models regrettably ignore the static interaction information among historical interactions and embedding dimensions.

In the meantime, there are several works showing the importance and effectiveness of modeling interaction information for recommendation tasks. The representative works in this field are FM and its extensions  \cite{rendle2010factorization,xin2019cfm,he2017neural}. The typical paradigm of FM-based methods is to model the second-order interaction between feature vectors. For example, NFM \cite{he2017neural} is proposed to model non-linear pairwise feature interactions. 
Although FM-based methods generally achieve satisfactory performance in recommendation tasks, 
existing works on modeling feature interaction mainly focus on context-aware recommendation tasks \cite{xin2019cfm,cheng2016wide, wang2018tem}. However, such contextual features are not always available, in particular, the user has very little historical data. 
Recently, some works focus on exploiting the dimension-level interaction information to enhance the performance of recommendation. For example, ConvNCF \cite{he2018outer} applies an outer-product operation to encode pairwise dimension-level interactions.
CFM \cite{xin2019cfm} is proposed to model second-order interactions for the context-aware recommendation.
In this work, we propose to model the high-order interaction among historical interactions and embedding dimensions simultaneously.
Instead of modeling interaction effects from contextual features like FM-based methods, we exploit interaction signals from implicit feedback data: the interacted users and items are treated as "contextual features" in our work. 
In this way, COMET captures the internal interaction patterns among the target user’s and target item’s historical interactions and the dimensional interaction signals among all latent dimensions, which has not been studied before. 
Moreover, we present how to enrich the representations of users and items by learnt interaction information, in this way, users' representations and items' representations are learnt dependently and lead to better recommendation performance.

\section{Preliminaries}
Before we detail our proposed method, we first formulate the problem and define the notations used in this paper.

\subsection{Problem Formulation and Notations}
Let $\mathcal{U}=\{u_{1},u_{2},\cdots, u_{m}\}$ be the set of users, and $\mathcal{I}=\{i_{1},i_{2},\cdots, i_{n}\}$ be the set of items. The user-item interaction matrix is denoted by $\mathbf{Y}=[y_{ui}]$ of size $m \times n$ from implicit feedback data as:
\begin{gather}
y_{ui}=\begin{cases}
1,&\text{if the interaction between $u$ and $i$ is observed}; \\
0,&\text{otherwise}.
\end{cases}
\end{gather}
Specifically, $y_{ui}=1$ represents the existence of observed interaction between user $u$ and item $i$, while $y_{ui}=0$ means the user-item interaction was not observed. Intuitively, the goal of recommendation is to compute the interaction scores of the missing entries in $\mathbf{Y}$, and a meaningful recommendation list can be further generated based on the estimated interaction scores.

Throughout this paper, we use $u$ and $i$ to represent a user and an item, respectively. We use bold symbols in lower case (e.g., $\mathbf{u}$) to denote vectors and bold symbols in upper case (e.g., $\mathbf{Y}$) to represent matrices. Moreover, $y_{ui}$ denotes the $(u,i)$-th element of matrix $\mathbf{Y}$. In addition, we denote predicted values by a \texttt{\char`\^} over it, for example, the final predicted interaction score between user $u$ and item $i$ is represented as $\hat{y}_{ui}$.

\subsection{Relationship with Matrix Factorization}
MF plays an important role in latent factor models, which factorizes the rating matrix into a user matrix and an item matrix for rating prediction \cite{koren2009matrix, NCF}. We denote the latent representations for user $u$ and item $i$ as $\mathbf{p}_u$ and $\mathbf{q}_i$, respectively. MF estimates the interaction score $\hat{y}_{ui}$ of $y_{ui}$ by the inner product of $\mathbf{p}_u$ and $\mathbf{q}_i$:
\begin{gather}
\hat{y}_{ui} = \mathbf{p}_u^\top\mathbf{q}_i=\sum_{k=1}^Kp_{uk}q_{ik}
\end{gather}
where $K$ denotes the dimension of the latent representations. Based on the above equation, MF linearly combines the latent features. 
While our proposed COMET model aims to enrich the users' and items' representations generated by MF. In COMET, a prediction $\hat{y}_{ui}$ of $y_{ui}$ as follows:
\begin{gather}
\hat{y}_{ui}=\sigma(\mathbf{h}^\top((\mathbf{p}_u+\mathbf{p}_u\prime)\odot(\mathbf{q}_i+\mathbf{q}_i\prime)))
\end{gather}
where $\odot$ represents the element-wise product between two vectors, $\sigma(\cdot)$ is the Sigmoid function, and $\mathbf{h}$ denotes a weight vector. 
If we set the weight vector $\mathbf{h}=\mathbf{1}$ where $\mathbf{1}$ is a vector whose elements are all equal to 1, and set the interaction vectors $\mathbf{p}_u\prime$ and $\mathbf{q}_i\prime$ = $\mathbf{0}$ where $\mathbf{0}$ is a vector whose elements are all equal to 0.
As can be seen, the MF model is exactly recovered by COMET except for the activation function, since $\mathbf{p}_u^\top\mathbf{q}_i=\mathbf{1}^\top((\mathbf{p}_u+ \mathbf{0}) \odot (\mathbf{q}_i+ \mathbf{0}))$. In the next section, we will introduce how to model the interaction vectors.

\section{The Proposed Approach}
\begin{figure*}[!t]
  \centering
  \includegraphics[width=0.75\linewidth]{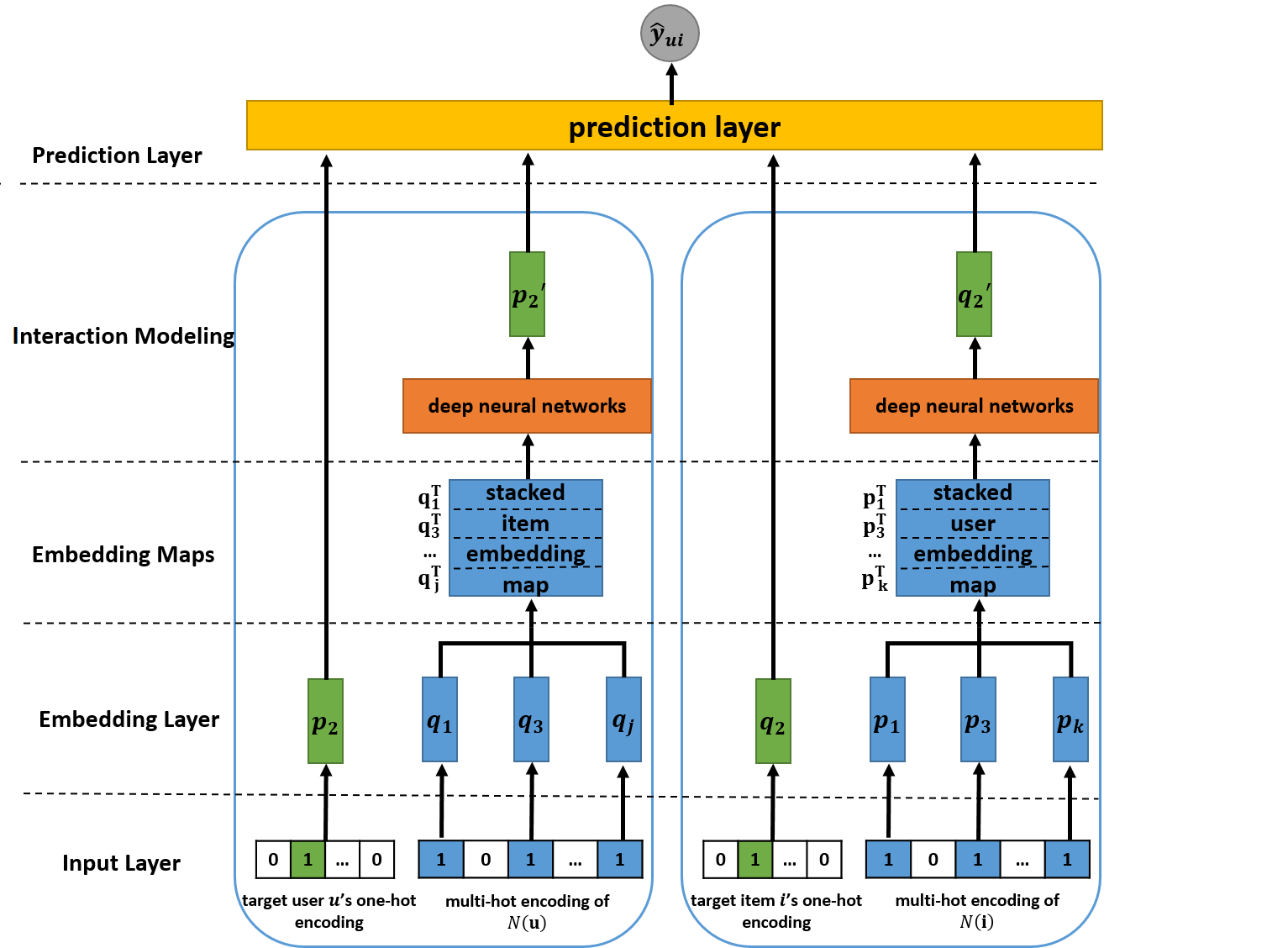}
  \caption{Framework of COMET (details of interaction modeling are given in Figure 2).}
\end{figure*}

Figure 1 demonstrates the proposed framework which encodes the high-order interactions among historical interactions and dimensional interactions to enrich the representations of the target user and the target item.
In this section, we detail our proposed COMET layer by layer.

\subsection{Input Layer.}
Most of the latent factor models \cite{he2017neural,he2018outer} only take one-hot encoding on the target user's ID and the target item's ID into account for the input layer. In this work, we also consider multi-hot encoding on the target user $u$'s interacted items as well as the target item $i$'s interacted users. 
Such a design will not only take into account more historical information but also benefit the construction of embedding maps in the next layer.

Let us take the target user $u$ and her interacted items $\text{N}(u)$ as an example. The one-hot encoding of $u$ can be presented as a vector $\mathbf{u}\in \{0,1\}^m$ whose entry indicates the user ID of the target user, and $m$ is the number of users. 
Similarly, the multi-hot encoding on interacted items $\text{N}(u)$ can be represented as a vector $\mathbf{u'}\in \{0,1\}^n$, the entries record the ID of items that the target user has interacted with, where $n$ denotes the number of items. In other words, the multi-hot encoding vector indicates the target user interacted with $j$-th item before if the $j$-th element of the multi-hot encoding vector $\mathbf{u}'_{j}$ is a non-zero value. 

\subsection{Embedding Layer.}
The embedding layer projects the target user $u$ and the target item $i$ to the latent space and gains the feature vectors $\mathbf{p}_u\in\mathbb{R}^K$ and $\mathbf{q}_i\in\mathbb{R}^K$, respectively. Similarly, we can obtain $\{\mathbf{q}_j\in\mathbb{R}^K|j\in\text{N}(u)\}$ and $\{\mathbf{p}_k\in\mathbb{R}^K|k\in\text{N}(i)\}$ for each interacted item $j\in\text{N}(u)$ and each interacted user $k\in\text{N}(i)$, where $K$ represents the embedding size. 
Note that the one-hot encoding vector and multi-hot encoding vector are fixed, because they only encode the user ID information and the interacted items' ID information. 
The embedding of the target user and interacted items shown in the embedding layer are learnt with the model in an end-to-end manner.

\subsection{Embedding Maps.} 
In this work, we treat the historical interactions as "contextual features". Their embeddings are horizontally stacked as "embedding maps" above the embedding layer.
For example, given a set of interacted items' embeddings for a user $\{\mathbf{q}_1, \mathbf{q}_3, ..., \mathbf{q}_j\}$, the stacked item embedding map is constructed as follows:
\begin{gather}\label{equ:comet_map}
\mathbf{E}_{i}^{|\text{N}(u)|\times K} = 
\begin{bmatrix}
\mathbf{q}_1^T\\
\mathbf{q}_3^T\\
...\\
\mathbf{q}_j^T
\end{bmatrix}
\end{gather}
In this way, the historical interactions are denoted as a matrix form.
Likewise, for a set of interacted users' embeddings $\{\mathbf{p}_1, \mathbf{p}_3, ..., \mathbf{p}_k\}$, the stacked user embedding map can be represented as:
\begin{gather}
\mathbf{E}_{u}^{|\text{N}(i)|\times K} = 
\begin{bmatrix}
\mathbf{p}_1^T\\
\mathbf{p}_3^T\\
...\\
\mathbf{p}_k^T
\end{bmatrix}
\end{gather}
where $\mathbf{E}_i$ and $\mathbf{E}_u$ represent the stacked item embedding map and stacked user embedding map, respectively. Note that $|\text{N}(i)|$ and  $|\text{N}(u)|$ represent the cardinality of $\text{N}(u)$ and $\text{N}(i)$, which are the number of historical interactions and the values can be controlled by historical data sampling.

Constructing such embedding maps is advantageous threefold. 
Firstly, by representing the historical interactions as embedding maps, our model is able to exploit the interaction signals internally (i.e. relationships among items and relationships among users), which empowers our model to learn users' representations and items' representations
Secondly, different from the outer-product operation which only considers pairwise dimensional interactions, our proposed embedding map reserve the latent information in the original embedding space, which enables our model to explicitly capture high-order dimensional interactions.
Thirdly, such a design of embedding maps makes it possible for our proposed COMET to capture the internal interactions and dimensional interactions simultaneously.
  
\begin{figure*}[!t]
  \centering
  \includegraphics[width=0.95\linewidth]{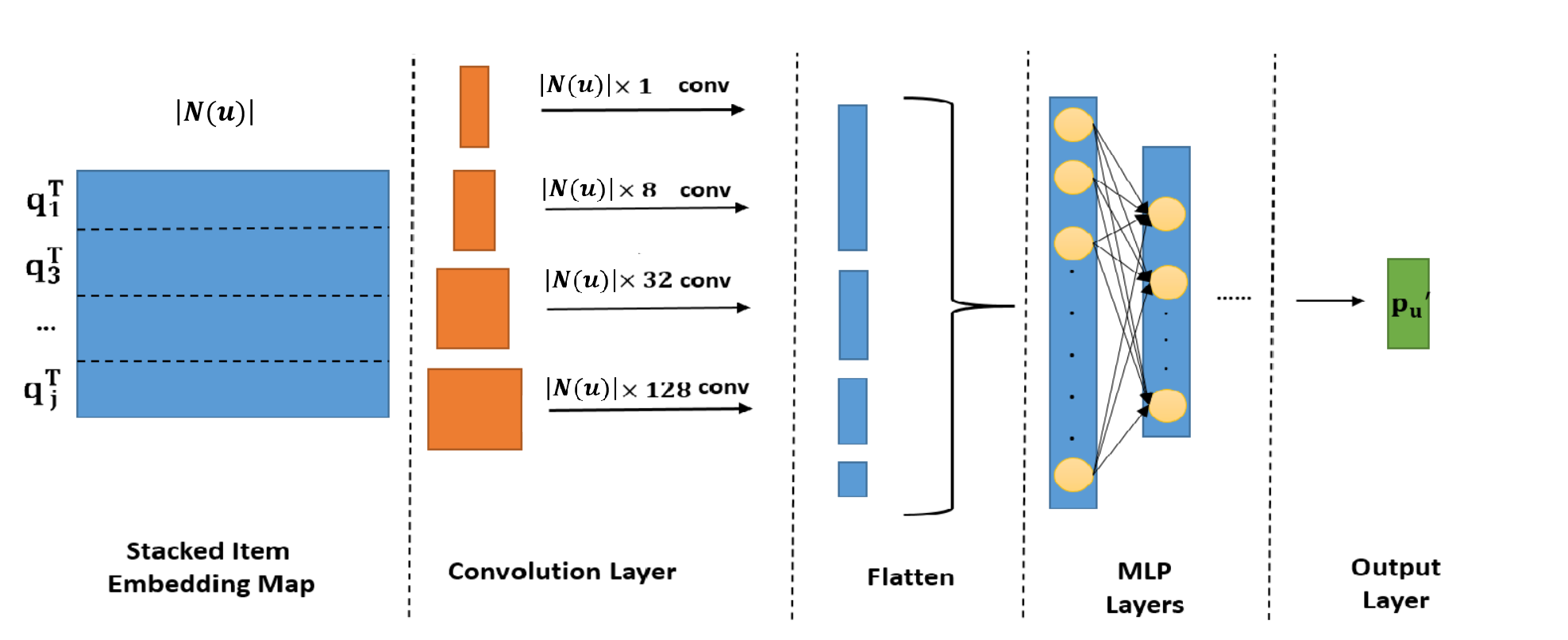}
  \caption{Illustration of the process of Interaction modeling.}
  \label{Inter_Model}
\end{figure*}
\subsection{Interaction Modeling.} 
The latent representations characterize both users and items by vectors of factors, and a high matching score between user and item factors leads to a recommendation \cite{koren2009matrix}. 
Therefore, modeling the dimensional interactions among such factors are important to achieve personalized recommendations.
In this subsection, since the user interaction vector and item interaction vector can be obtained from the same process, we focus on illustrating how to obtain the user interaction vector. 
As shown in Figure \ref{Inter_Model}, the interaction modeling process aims to generate an interaction vector which encodes the high-order interaction information among the item embedding map constructed in the previous layer.
Technically speaking, any method that can transform a matrix into a vector can be used here.
Intuitively, MLP is a common choice to capture high-order interactions, which has been widely used in recommendation research \cite{he2017neural, NCF,hornik1991approximation}. However, recent studies \cite{he2018outer,xin2019cfm} demonstrate that the interactions are inefficiently and implicitly modeled by MLP with current optimization techniques, resulting in sub-optimal performance on recommendation tasks.
Inspired by the recent works that explicitly encode the pairwise interactions and treat the pairwise interaction map as a 2D image or 3D cube \cite{he2018outer, xin2019cfm, han2019convolutional,yan2019cosrec}. 
We propose to use a single-layer CNN with filters of variant sizes to capture the high-order interactions encoded in the item embedding map. 
A fully-connected MLP is then used to generate the user interaction vector. 
The efficacy of such a design is studied in Section 4.2.
Specifically, the cross features generated by the convolution of the item embedding map $\mathbf{E}_i$ and $l-th$ filter are denoted as: 
\begin{gather}
\mathbf{c}_i^l=\psi (\mathbf{W}_i^l * \mathbf{E}_i + \mathbf{b}_i^l)
\end{gather}
where $*$ represents a convolution operator, $\mathbf{W}_i^l\in\mathbb{R}^{|\text{N}(i)| \times H}$ is a weight matrix and $\mathbf{b}_i^l$ is the corresponding bias.
Besides, $\psi(\cdot)$ is a non-linear activation function, here we employ the rectified linear unit (ReLU) \cite{yin2021iav,yin2021virprenet,nair2010rectified}.
Note that $|\text{N}(i)| \times H$ denotes the size of the filter, aiming to cover all the rows (interacted items) of the embedding map.
We use multiple filters with varying widths such as \cite{kim2014convolutional} to extract features from both the local and global scale. 
\begin{gather}
\mathbf{c}_i= [ \mathbf{c}_i^1 ; \mathbf{c}_i^2 ; ... ; \mathbf{c}_i^l ],
\end{gather}
where $\mathbf{c}_i$ represents the item internal interaction features.

Above the CNN is a fully-connected MLP layer, it takes $\mathbf{c}_i$ as input and generates the user interaction vector.
\begin{equation}
\begin{split}
&\mathbf{p}_u^1= \psi_2 (\mathbf{W}_i^1 \mathbf{c}_i + \mathbf{b}_i^1)\\
&\mathbf{p}_u^2= \psi_2 (\mathbf{W}_i^2 \mathbf{p}_u^1 + \mathbf{b}_i^2)\\
&...\\
&\mathbf{p}_u^L= \psi_2 (\mathbf{W}_i^L \mathbf{p}_u^{L-1} + \mathbf{b}_i^L)\\
\end{split}
\end{equation}
where the number of hidden layers is denoted by $L$, $\mathbf{W}_i^L$ represents the weight matrix, $\mathbf{b}_i^L$ is a bias vector, and $\psi_2$ means the activation function for MLP layers. 
The output of the last hidden layer $\mathbf{p}_u^L$ is then transformed to the user interaction vector $\mathbf{p}_u\prime$:
\begin{gather}\label{equ:comet_neibor_vec}
\mathbf{p}_u\prime= \mathbf{W}_i \mathbf{p}_u^L + \mathbf{b}_i
\end{gather}
where $\mathbf{W}_i$, $\mathbf{b}_i$ represent the weight matrix and bias vector for the output layer.
We can obtain the item interaction vector $\mathbf{q}_i\prime$ in the same way.
  
\subsection{Prediction Layer.} 
Given two interaction vectors $\mathbf{p}\prime_u$ and $\mathbf{q}\prime_i$ for $\mathbf{p}_u$ and $\mathbf{q}_i$, the original representations and the learnt interaction information are combined in the prediction layer. 
To be specific, the predicted interaction score between the target user and the target item is predicted as follows:
\begin{gather}\label{comet_prediction}
\hat{y}_{ui}=\sigma(\mathbf{h}^\top((\mathbf{p}_u+\mathbf{p}_u\prime)\odot(\mathbf{q}_i+\mathbf{q}_i\prime)))
\end{gather}
where $\mathbf{h}$ denotes a learnt weight vector, the sigmoid function is used as the activation function, and $"\odot"$ represents the element-wise product between two vectors.
By enriching the original representations of the target user and target item with internal interaction vectors, users' representations and items' representations are dependently learnt. The efficacy of such a design is discussed in Section 4.2.

\subsection{Loss Function.} 
In this paper, we focus on the task of recommendation from implicit feedback data. To this end, our model should learn parameters with a ranking-aware objective. Therefore, Binary Cross Entropy (BCE) loss which constrains the output in the range of $[0,1]$ is employed:

\begin{align}\label{equ:COMET_loss}
\mathcal{L}=-\sum_{(u,i)\in\mathcal{O}^{+}\cup\mathcal{O}^{-}}y_{ui}\log\hat{y}_{ui}+(1-y_{ui})\log(1-\hat{y}_{ui})
\end{align}
where $\mathcal{O}^+$ is the set of positive samples and $\mathcal{O}^-$ represents the set of negative samples, respectively. During the training process, four negative samples are randomly sampled for each positive sample in every single training epoch.

\begin{algorithm}[t]
\caption{The COMET Algorithm} \label{alg1}

\begin{algorithmic}[1]
\State \textbf{Input}: Interaction matrix Y, number of neighbors $|\text{N}(u)|$ and $|\text{N}(u)|$
\State \textbf{Output}: Model parameters $\theta$
\For{each $(u,i)\in\mathcal{O}^+ $}
\State Draw negative instances $(u,j)\in\mathcal{O}^-$;
\EndFor

\For {each $(u,i) \in$ a mini-batch}
 \State Represent each user $u$ as $\mathbf{p}_u$, and each item $i$ as $\mathbf{q}_i$; 
  \State Collect historical interactions $\text{N}(u)$ and $\text{N}(i)$; 
  \State Compute $\mathbf{p}_u\prime$ and $\mathbf{q}_i\prime$ with Eqs.~(\ref{equ:comet_map}-\ref{equ:comet_neibor_vec});
 \State Compute $\hat{y}_{ui}$ based on Eq. \ref{comet_prediction};
  \State Calculate $\mathcal{L}$ based on  Eq. \ref{equ:COMET_loss};
  \State Update $\theta$ to minimize $\mathcal{L}$, using $\nabla_{\theta}\mathcal{L}$
\EndFor
\end{algorithmic}

\end{algorithm}

\section{Experiments}
We aim to evaluate the effectiveness and rationality of our proposed method in this section. We hence design extensive experiments and ablation studies in order to answer the following research questions:
\begin{itemize}
\item\textbf{RQ1} Is COMET able to outperform the state-of-the-art latent factor models?
\item\textbf{RQ2} Can our method effectively capture the interaction information from historical interactions and embedding dimensions?
\item\textbf{RQ3} Does COMET benefit from the learnt internal interaction signals?
\item\textbf{RQ4} How do the key hyperparameters influence the performance of our method?
\end{itemize}

\subsection{Experimental Settings}
\subsection{Datasets.} 
We conduct experiments and sensitivity analysis on three public datasets: Amazon Movies \& Tv \footnote{Amazon Movies and Tv: \url{http://jmcauley.ucsd.edu/data/amazon/}}, and Amazon CDs \& Vinyl,
 and MovieLens 1M (ML-1M) dataset\footnote{MovieLens 1M: \url{https://grouplens.org/datasets/movielens/}},
Since it is difficult to evaluate recommendation models on a highly sparse dataset, we follow the common practice \cite{rendle2009bpr, NCF}, ignoring the users with less than 10 interactions for both Amazon datasets.
Noted that COMET aims to generate personalized recommendation from implicit feedback, we hence convert all ratings to implicit feedback to indicate whether the user has interacted with the item, by representing the rating entries as either 1 or 0. The characteristics of the three datasets are shown in Table 1.

\begin{table*}[!t]
\centering\caption{Characteristics of datasets.}
\resizebox{0.8\textwidth}{!}{
\setlength{\tabcolsep}{6mm}{
\begin{tabular}{c|c|c|c}
\hline
\hline
Dataset & ML-1M &  Movies\&Tv & CDs\&Vinyl \\
\hline
Number of users & 6,040  & 40,928 & 26,876 \\
\hline
Number of Items & 3,706  & 51,509 & 66,820\\
\hline
Number of interactions & 1,000,209  & 1,163,413 & 770,188\\
\hline
Rating density & 0.04468  & 0.00055 & 0.00043\\
\hline
\hline
\end{tabular}
}
}
\label{tab1}
\end{table*}

\begin{table*}[!t]
\centering\caption{HR@5 and NDCG@5 of all recommendation models are evaluated. The best results are highlighted. In addition,  $\sbullet[1.5]/\Circle$ indicates whether the performance of COMET is significantly superior to the compared methods on each dataset. (Paired t-test at
0.05 significance level)}
\resizebox{0.9\textwidth}{!}{
\setlength{\tabcolsep}{4mm}{
\begin{tabular}{c|cc|cc|cc}
\hline\hline
    \multicolumn{1}{c|}{\multirow{2}{*}{Method}}  & \multicolumn{2}{c|}{ML-1M} & \multicolumn{2}{c|}{Movie\&Tv} & \multicolumn{2}{c}{CDs\&Vinyl}  \\ \cline{2-7}
       & HR@5        & NDCG@5            & HR@5          & NDCG@5         & HR@5           & NDCG@5         \\ \hline
BCE-MF & 0.540$\pm 0.001$ •         & 0.376$\pm 0.002$ •                & 0.634$\pm 0.002$            & \textbf{0.498}$\pm \textbf{0.001}$             & 0.606$\pm 0.001$ •            & 0.466$\pm 0.001$ •           \\ \hline
SVD++  & 0.557$\pm 0.002$      & 0.388$\pm 0.001$ •  &0.606$\pm 0.001$ •         & 0.462$\pm 0.001$ •        & 0.607$\pm 0.002$ •         & 0.466$\pm 0.001$ •          \\ \hline
FISM   & 0.528$\pm 0.002$ •      & 0.372$\pm 0.002$ •    & 0.583$\pm 0.003$ •       & 0.452$\pm 0.002$ •         & 0.592$\pm 0.001$ •         & 0.457$\pm 0.002$ •          \\ \hline
MLP    & 0.526$\pm 0.003$ •        & 0.362$\pm 0.003$ •               & 0.570$\pm 0.002$ •         & 0.425$\pm 0.001$ •          & 0.588$\pm 0.003$ •          & 0.445$\pm 0.001$ •          \\ \hline
GMF    & 0.540$\pm 0.001$ •      & 0.372$\pm 0.001$ •             & 0.569$\pm 0.004$ •         & 0.427$\pm 0.003$ •          & 0.620$\pm 0.004$ •         & 0.481$\pm 0.002$ •          \\ \hline
NeuMF  & 0.548$\pm 0.003$ •      & 0.381$\pm 0.002$ •               & 0.596$\pm 0.001$ •         & 0.453$\pm 0.001$ •          & 0.629$\pm 0.001$ •         & 0.491$\pm 0.001$ •         \\ \hline
ConvNCF& 0.539$\pm 0.001$ •     & 0.376$\pm 0.001$ •              & 0.623$\pm 0.002$ •         & 0.488$\pm 0.001$ •          & 0.603$\pm 0.002$ •          & 0.457$\pm 0.001$ •          \\ \hline
LightGCN& 0.542$\pm 0.002$ •          & 0.381$\pm 0.001$ •                 & 0.618$\pm 0.001$ •           & 0.476$\pm 0.001$ •            & 0.658$\pm 0.004$ •         & 0.525$\pm 0.003$             \\ \hline
COMET   & \textbf{0.558$\pm \textbf{0.001}$ }       & \textbf{0.392$\pm \textbf{0.001}$ }               & \textbf{0.637$\pm \textbf{0.002}$ }         & 0.491$\pm 0.002$            & \textbf{0.667$\pm \textbf{0.002}$  }          & \textbf{0.528$\pm \textbf{0.002}$  }          \\ \hline\hline
\end{tabular}
}
}
\end{table*}

\begin{table*}[!t]
\centering\caption{HR@10 and NDCG@10 of all recommendation models are evaluated. The best results are highlighted. In addition,  $\sbullet[1.5]/\Circle$ indicates whether the performance of COMET is significantly superior to the compared methods on each dataset. (Paired t-test at 0.05 significance level)}
\resizebox{0.9\textwidth}{!}{
\setlength{\tabcolsep}{4mm}{
\begin{tabular}{c|cc|cc|cc}
\hline\hline
   \multicolumn{1}{c|}{\multirow{2}{*}{Method}}   & \multicolumn{2}{c|}{ML-1M} &  \multicolumn{2}{c|}{Movie\&Tv} & \multicolumn{2}{c}{CDs\&Vinyl}  \\ \cline{2-7}
       & HR@10       & NDCG@10     & HR@10         & NDCG@10        & HR@10          & NDCG@10 \\ \hline
BCE-MF & 0.706$\pm 0.001$ •        & 0.429$\pm 0.002$ •               & 0.738$\pm 0.002$ •           & \textbf{0.533$\pm \textbf{0.001}$  }          & 0.727$\pm 0.001$ •           & 0.509$\pm 0.001$ •           \\ \hline
SVD++  & 0.713$\pm 0.003$ •      & 0.438$\pm 0.002$ •              &0.728$\pm 0.001$ •          &0.502$\pm 0.001$ •          & 0.719$\pm 0.001$ •         & 0.502$\pm 0.001$ •         \\ \hline
FISM   & 0.699$\pm 0.003$ •      & 0.433$\pm 0.001$ •               & 0.708$\pm 0.002$ •         & 0.471$\pm 0.001$ •          & 0.729$\pm 0.002$ •         & 0.512$\pm 0.001$ •        \\ \hline
MLP    & 0.703$\pm 0.003$ •      & 0.421$\pm 0.004$ •              & 0.703$\pm 0.002$ •         & 0.471$\pm 0.001$ •         & 0.712$\pm 0.003$ •         & 0.485$\pm 0.004$ •          \\ \hline
GMF    & 0.711$\pm 0.001$ •      & 0.429$\pm 0.002$ •              & 0.712$\pm 0.006$ •         & 0.479$\pm 0.005$ •         & 0.729$\pm 0.006$ •          & 0.515$\pm 0.002$ •         \\ \hline
NeuMF  & 0.727$\pm 0.004$       & 0.443$\pm 0.002$ •              & 0.721$\pm 0.001$ •        & 0.493$\pm 0.002$ •          & 0.750$\pm 0.001$ •         & 0.529$\pm 0.001$ •         \\ \hline
ConvNCF& 0.710$\pm 0.002$ •      & 0.431$\pm 0.001$ •              & 0.726$\pm 0.001$ •         & 0.521$\pm 0.001$ •          & 0.722$\pm 0.001$ •          & 0.496$\pm 0.001$ •         \\ \hline
LightGCN& 0.709$\pm 0.003$ •         & 0.434$\pm 0.002$ •                & 0.744$\pm 0.002$ •            & 0.517$\pm 0.001$ •             & 0.757$\pm 0.003$ •            & 0.556$\pm 0.003$             \\ \hline
COMET   & \textbf{0.729$\pm \textbf{0.002}$ }       & \textbf{0.448$\pm \textbf{0.001}$ }    & \textbf{0.759$\pm \textbf{0.002}$ }         & 0.529$\pm 0.003$           & \textbf{0.780$\pm \textbf{0.003}$ }          & \textbf{0.560$ \pm \textbf{0.004}$ }          \\ \hline\hline
\end{tabular}
}
}
\end{table*}
\subsection{Compared Methods.} 
To demonstrate the effectiveness of COMET, we also study the performance of the following state-of-the-art counterparts:
\begin{itemize}
\item\textbf{MF}\cite{rendle2009bpr} It is the classic MF trained by optimizing the binary cross entropy loss.
\item\textbf{SVD++}\cite{svd} SVD++ enriches the user latent factor with her interacted items' embedding.
\item\textbf{FISM}\cite{fism} As an item-based latent factor model, FISM factorizes the item-item similarity matrix into two low-dimensional latent factor matrices.
\item\textbf{GMF}\cite{NCF} GMF generalized the MF model in a non-linear manner.
\item\textbf{MLP}\cite{NCF} The interaction function between users' and items' representations is learnt by a MLP.
\item\textbf{NeuMF}\cite{NCF} NeuMF combines of GMF and MLP. We compare with NeuMF-p which pre-trains GMF and MLP\cite{NCF}.
\item\textbf{ConvNCF}\cite{he2018outer} It uses outer-product to model the pairwise interactions between the latent dimensions. A CNN is then used to discover the high-level interactions among embedding dimensions. 
\item\textbf{LightGCN}\cite{he2020lightgcn} A state-of-the-art Graph Convolution Network-based recommendation approach.

\end{itemize}
\subsection{Training Details.} 
In order to find out the optimal parameter settings for the comparing approaches, we carefully tune hyperparameters suggested by the respective literature. 
To be specific, for all the recommendation models, we choose the learning rate from $[5\text{e}^{-7},1\text{e}^{-6},5\text{e}^{-6},1\text{e}^{-5},5\text{e}^{-5}, 1\text{e}^{-4}, 5\text{e}^{-4}, 1\text{e}^{-3}, 5\text{e}^{-3}]$, we select the embedding size $K$ from the following set: $[16, 32, 64, 128]$, and the regularization parameter tried lies in the
interval $[5\text{e}^{-8},1\text{e}^{-7},5\text{e}^{-7},1\text{e}^{-6}, 5\text{e}^{-6}, 1\text{e}^{-5}, 5\text{e}^{-5}]$.
Since there are multiple fully-connected layers in the MLP and NeuMF \footnote{\url{https://github.com/hexiangnan/neural_collaborative_filtering}}, the number of hidden layers has been fairly tuned from 1 to 3 \cite{NCF}. 
As for ConvNCF\footnote{\url{https://github.com/duxy-me/ConvNCF}} and LightGCN \footnote{\url{https://github.com/wubinzzu/NeuRec}} \cite{he2020lightgcn}, we follow the settings proposed in \cite{he2018outer}. 
Note that we trained MF and ConvNCF with binary cross entropy loss like \cite{rendle2020neural}, so as to conduct a fair comparison among all baselines and our proposed method. In addition, all the recommendation approaches are trained until convergence.

For our proposed method, the weight vectors are initialized by the Xavier initialization \cite{glorot2010understanding}. Moreover, we initialize the embedding layer and weight matrices for CNN by the uniform distribution.
In addition, we employ the Adaptive Moment Estimation optimizer (Adam) \cite{kingma2014adam} to train our proposed model, and implement our proposed method using PyTorch\cite{paszke2017automatic}. 
The learning rate we tried are $[5\text{e}^{-5}, 1\text{e}^{-4}, 5\text{e}^{-4}, 1\text{e}^{-3}, 5\text{e}^{-3}]$, and the regularization parameter we tried are: $[1\text{e}^{-6}, 5\text{e}^{-6}, 1\text{e}^{-5}, 5\text{e}^{-5}]$.
Moreover, the embedding size $K$ is fixed at 128 and the dropout rate at 0.3 which always achieves better results under our setting. As for CNN, we empirically set the number of channels, stride, and padding to 8, 1, and 0, respectively. The filters of CNN are designed to cross all the latent dimensions, for example, the sizes of filters for the item embedding map are $|\text{N}(u)| \times 1$, $|\text{N}(u)| \times 8$, $|\text{N}(u)| \times 32$, $|\text{N}(u)| \times 128$, respectively. 
However, a user may have interacted with many items and leading to a large $|\text{N}(u)|$ in the real-world scenario and thus may need intensive computational power. To alleviate the problem, we empirically set the maximum number of interactions to 50 in the experiment.

\subsection{Experimental Results}
\subsection{Evaluation Protocols.} 
In order to make a fair comparison among COMET with other approaches, the \textsl{leave-one-out} evaluation method is adopted, which is the common choice for recommendation from implicit feedback \cite{rendle2009bpr,he2018outer}. 
To be specific, we randomly sampled one interaction for each user as the validation set, on which we tune hyper-parameters of all approaches.
Then we hold the latest historical item of each user as the test positive samples and the other 99 random items which have no interaction with this user as the test negative samples. Therefore, all the comparing models generate recommendations for each user by ranking the above 100 mentioned items.

To evaluate the quality of the generated recommendation list, we employ two evaluation metrics in this paper, namely \textsl{Hit Ratio} (HR) and \textsl{Normalized Discounted Cumulative Gain} (NDCG). HR@$k$ measures if the testing item is included in the top-$k$ recommendation list, and NDCG@$k$ takes the position of correct recommendations into account \cite{he2018outer}.

\subsection{Performance Comparison (The Answer to RQ1)}
Table 2 and Table 3 show the top-k evaluation on all three datasets. We run 5 times for each method and perform the significant test.
Obviously, COMET achieves the best performance on all datasets regarding both HR and NDCG. 
We believe that the underlying factor is the interaction modeling process. By efficiently capturing the high-order interaction information among embedding maps, better representations of users and items are modeled.
Besides, we can see that SVD++ achieves comparable performance to some deep models (i.e. MLP and ConvNCF) on the ML-1M dataset, which may benefit from the abundant latent factors of interacted items in the ML-1M dataset.

\begin{figure}[!t]
\centering
      \subfigure[HR@10]{
    \begin{minipage}[b]{0.37\textwidth}
      \centering
      \includegraphics[width=2in]{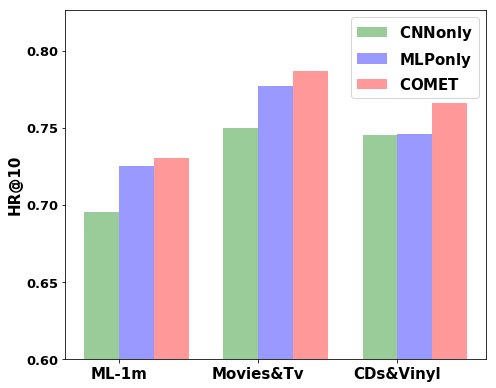}
    \end{minipage}}
      \subfigure[NDCG@10]{
    \begin{minipage}[b]{0.37\textwidth}
      \centering
      \includegraphics[width=2in]{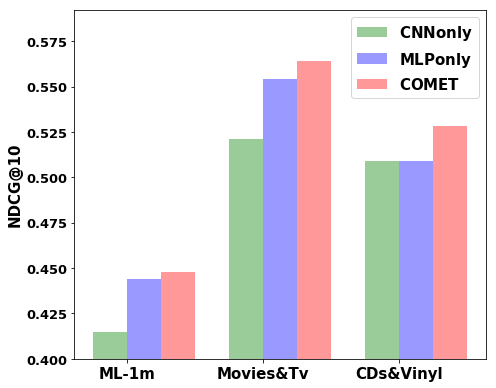}
    \end{minipage}}
    \vspace{-0.1in}
  \caption{Performance of different structures for interaction modeling on all the datasets.}
\end{figure}

\begin{table*}[!t]
\centering
\caption{Performance of COMET with different filter sets on the three datasets.}
\begin{tabular}{c|c|c|c|c|c}
\hline
\hline
 Datasets & Methods  & HR@5  & NDCG@5 & HR@10 & NDCG@10 \\ \hline
\multicolumn{1}{c|}{\multirow{4}{*}{ML-1M}} &COMET(1)          & 0.557 & 0.389  & \bf{0.730} & 0.445   \\ \cline{2-6}
\multicolumn{1}{c|}{} &COMET(1,8)        & 0.549 & 0.386  & 0.721 & 0.442   \\ \cline{2-6}
\multicolumn{1}{c|}{} &COMET(1,8,32)     & 0.553 & 0.385  & 0.722 & 0.440   \\ \cline{2-6}
\multicolumn{1}{c|}{} &COMET & \bf{0.558} & \bf{0.392}  & {0.729} & \bf{0.448}   \\ \hline\hline
\multicolumn{1}{c|}{\multirow{4}{*}{Movie\&Tv}}   &COMET(1)          & 0.631 & 0.483  & {0.758} & 0.524   \\ \cline{2-6}
\multicolumn{1}{c|}{} &COMET(1,8)        & 0.619 & 0.476  & 0.748 & 0.515   \\ \cline{2-6}
\multicolumn{1}{c|}{} &COMET(1,8,32)     & 0.619 & 0.477  & 0.749 & 0.517   \\ \cline{2-6}
\multicolumn{1}{c|}{} &COMET & \bf{0.637} & \bf{0.491}  & \bf{0.759} & \bf{0.529}   \\ \hline\hline
\multicolumn{1}{c|}{\multirow{4}{*}{CD\&Vinyl}}   &COMET(1)          & 0.642 & 0.513  & 0.749 & 0.547   \\ \cline{2-6}
\multicolumn{1}{c|}{} &COMET(1,8)        & 0.660 & 0.525  & 0.772 & 0.558   \\ \cline{2-6}
\multicolumn{1}{c|}{} &COMET(1,8,32)     & 0.665 & 0.523  & {0.779} & 0.559   \\ \cline{2-6}
\multicolumn{1}{c|}{} &COMET & \bf{0.667} & \bf{0.528}  & \bf{0.780} & \bf{0.560}   \\ \hline\hline
\end{tabular}
\end{table*}

\subsection{Study of interaction modeling (The Answer to RQ2)}
COMET aims to model the high-order interactions from historical interactions and embedding dimensions. 
In this work, we apply a single-layer CNN to extract cross features over the embedding map. With filters of variant sizes, interaction features can be effectively obtained on local and global scales. Those features are then served as the input of the fully-connected MLP layer, in this way, dimensional interaction signals are captured in a rather explicit manner. To clearly demonstrate the rationality of our proposed interaction modeling process, we present two models here:
\begin{itemize}
\item\textbf{COMET-CNN only} This method uses a 3-layer CNN to transform embedding maps to interaction vectors. An output layer is used to guarantee the dimension of the interaction vector to $K$. The filter size, number of channels, stride, and padding are set to 3$\times$3, 8, 2, and 0, respectively. 
\item\textbf{COMET-MLP only} Embeddings of the interacted users or the interacted items are concatenated and fed into a fully-connected MLP directly. Besides the dropout rate, we carefully tune the number of hidden layers from 1 to 3 according to the tower structure of neural networks \cite{NCF}.  
\end{itemize}

\begin{figure}[!t]
\centering
      \subfigure{
    \begin{minipage}[b]{0.3\textwidth}
      \centering
      \includegraphics[width=2in]{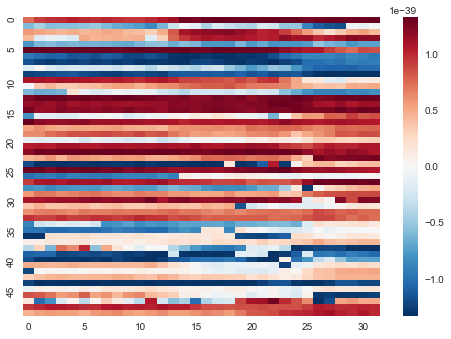}
    \end{minipage}}\hspace{10mm}
      \subfigure{
    \begin{minipage}[b]{0.3\textwidth}
      \centering
      \includegraphics[width=2in]{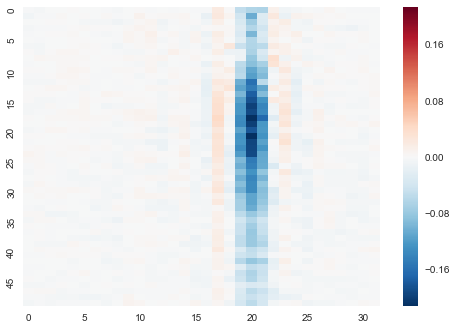}
    \end{minipage}}
  \caption{Visualisation of the kernels of CNN. The patterns differ significantly across different channels, hence different interaction signals are generated. As for the heat map, the vertical and horizontal axes represent the height and width of kernels, respectively. Meanwhile, the color represents the magnitude of its weight value. A color bias towards red indicates a larger weight value, while a bias towards blue indicates a smaller weight value.} Best viewed in color.\label{heat}
\end{figure}

The comparison among COMET, COMET-CNN only, and COMET-MLP only is displayed in figure 3. We can see that MLP is able to capture complex relationships better by encoding the high-order feature interactions with CNN in a rather explicit way. 
This observation agrees with the conclusion of recent works \cite{he2018outer,xin2019cfm}. 
Furthermore, to discover the benefit of modeling high-order interactions among all the latent dimensions, a fair comparison among different sets of filters is studied here. For example, COMET(1) means that only a filter $\mathbf{W}\in\mathbb{R}^{|\text{N}(i)| \times 1}$ is used and no dimensional interaction is captured. While COMET (i.e. COMET(1,8,32,128)) captures not only the independent dimensional information but also the interaction signals among all the embedding dimensions. 
The performance of COMET with different filter sets is shown in Table 4. 
We can see that COMET generally performs better than other counterparts, in particular under the NDCG evaluation metrics. This demonstrates the effectiveness of modeling interaction effects across latent dimensions.
In addition, we observe that the performance gaps between the methods are not as significant as expected.
The underlying reason may be the relatively small amount of input which encodes the interaction among all (i.e, K) the embedding dimensions. 
We may alleviate this problem by weighting or selecting the cross features before feeding them onto MLP. We leave this challenge as future work. 
Figure \ref{heat} shows the heat map of two randomly selected kernels. A quick observation is that the selected two kernels are able to capture different interaction signals among latent dimensions.

\begin{figure*}[t]
\centering
      \subfigure[HR@10]{
    \begin{minipage}[b]{0.37\textwidth}
      \centering
      \includegraphics[width=2in]{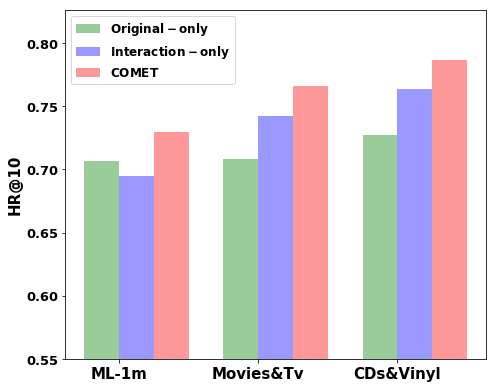}
    \end{minipage}}
      \subfigure[NDCG@10]{
    \begin{minipage}[b]{0.37\textwidth}
      \centering
      \includegraphics[width=2in]{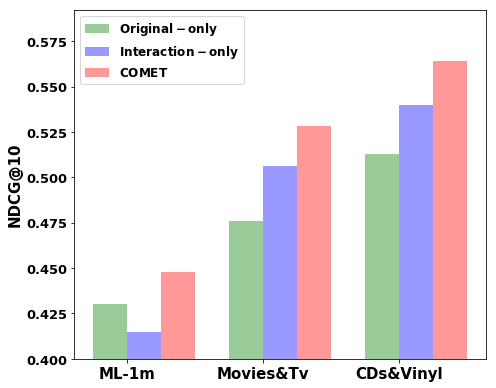}
    \end{minipage}}
  \caption{Comparison among COMET, COMET-original only, and COMET-interaction only on all used datasets.}
\end{figure*}

\begin{figure*}[!ht]
\centering

    \hspace{-0.1in}
    \begin{minipage}[b]{0.32\textwidth}
      \centering
      \includegraphics[width=1.65in]{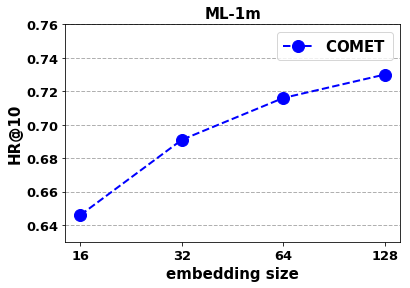}
    \end{minipage}%
        \hspace{0.05in}
    \begin{minipage}[b]{0.32\textwidth}
      \centering
      \includegraphics[width=1.65in]{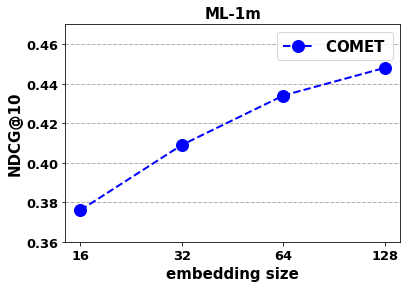}
    \end{minipage}
    \begin{minipage}[b]{0.32\textwidth}
      \centering
      \includegraphics[width=1.65in]{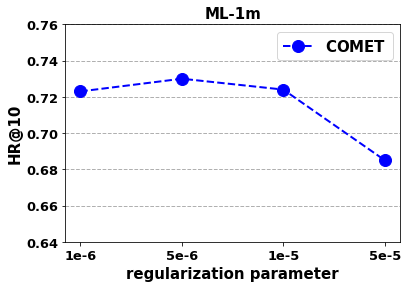}
    \end{minipage}
    \begin{minipage}[b]{0.32\textwidth}
      \centering
      \includegraphics[width=1.65in]{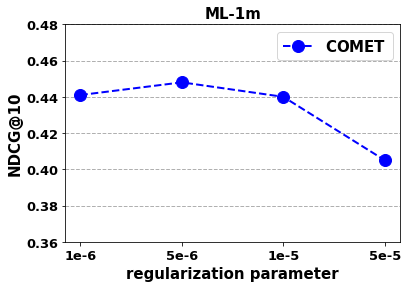}
    \end{minipage}
    \begin{minipage}[b]{0.32\textwidth}
      \centering
      \includegraphics[width=1.65in]{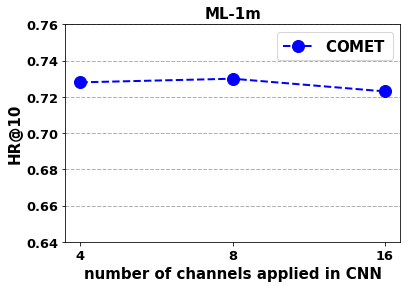}
    \end{minipage}
    \begin{minipage}[b]{0.32\textwidth}
      \centering
      \includegraphics[width=1.65in]{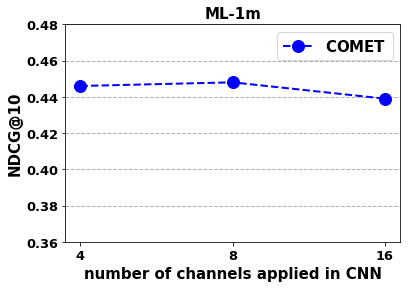}
    \end{minipage}

    \hspace{-0.1in}
    \begin{minipage}[b]{0.32\textwidth}
      \centering
      \includegraphics[width=1.65in]{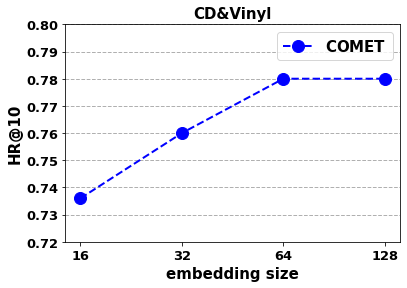}
    \end{minipage}
        \hspace{0.05in}
    \begin{minipage}[b]{0.32\textwidth}
      \centering
      \includegraphics[width=1.65in]{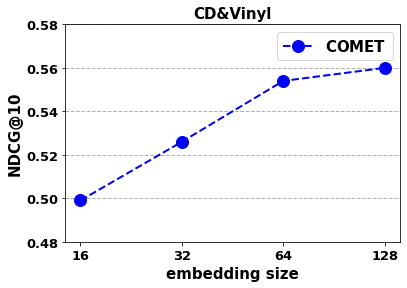}
    \end{minipage}
    \begin{minipage}[b]{0.32\textwidth}
      \centering
      \includegraphics[width=1.65in]{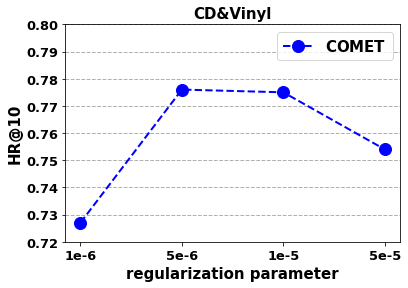}
    \end{minipage}
    \begin{minipage}[b]{0.32\textwidth}
      \centering
      \includegraphics[width=1.65in]{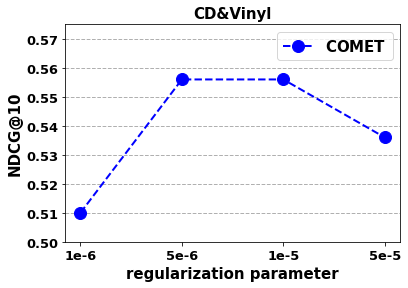}
    \end{minipage}
    \begin{minipage}[b]{0.32\textwidth}
      \centering
      \includegraphics[width=1.65in]{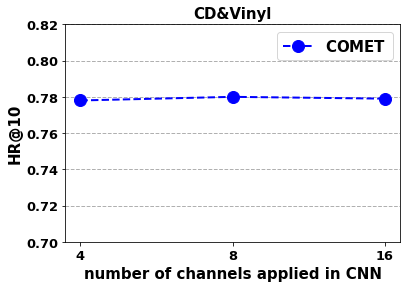}
    \end{minipage}
    \begin{minipage}[b]{0.32\textwidth}
      \centering
      \includegraphics[width=1.65in]{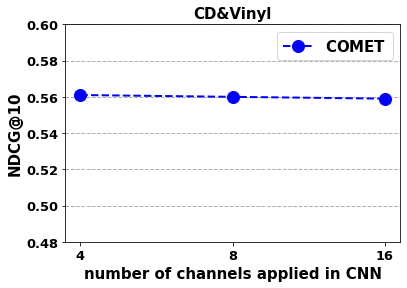}
    \end{minipage}

    \hspace{-0.1in}
    \begin{minipage}[b]{0.32\textwidth}
      \centering
      \includegraphics[width=1.65in]{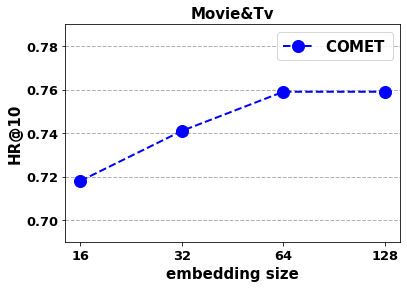}
    \end{minipage}
        \hspace{0.05in}
    \begin{minipage}[b]{0.32\textwidth}
      \centering
      \includegraphics[width=1.65in]{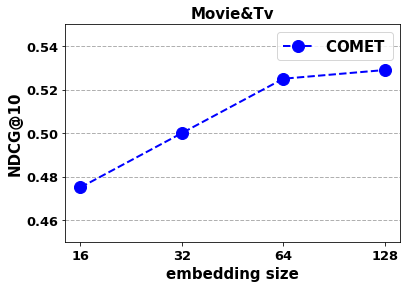}
    \end{minipage}
    \begin{minipage}[b]{0.32\textwidth}
      \centering
      \includegraphics[width=1.65in]{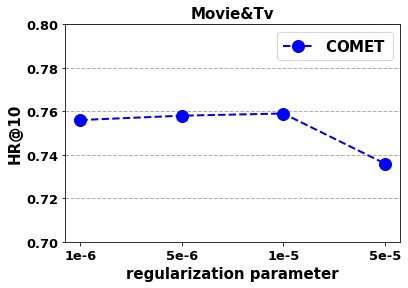}
    \end{minipage}
    \begin{minipage}[b]{0.32\textwidth}
      \centering
      \includegraphics[width=1.65in]{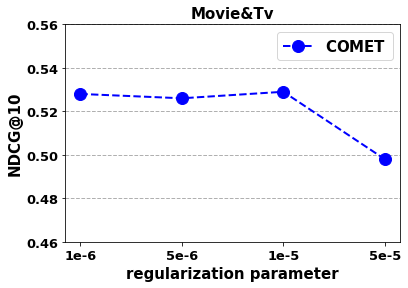}
    \end{minipage}
    \begin{minipage}[b]{0.32\textwidth}
      \centering
      \includegraphics[width=1.65in]{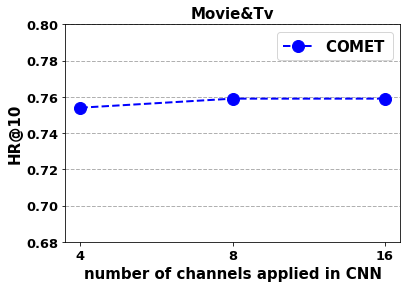}
    \end{minipage}
    \begin{minipage}[b]{0.32\textwidth}
      \centering
      \includegraphics[width=1.65in]{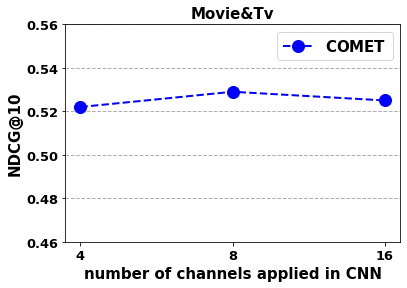}
    \end{minipage}
  \caption{Sensitivity analysis of COMET on all datasets.}
  
\end{figure*}

\subsection{Study of the interaction-aware representation (The Answer to RQ3)}

As mentioned before, the representations of users and items are enriched by learnt interaction vectors in the prediction layer. In order to study the efficacy of such a design, we design the following models:
\begin{itemize}
\item\textbf{COMET-original only} The interaction score is only predicted by the inner product of the original representations of the target user and target item. In this way, the step of constructing embedding maps and the interaction modeling process is omitted. 
\item\textbf{COMET-interaction only} Similarly, we ignore the original representations of the target user (i.e. $\mathbf{p}_u$) and the target item (i.e. $\mathbf{q}_i$) in the prediction layer in this approach. In other words, we only use $\mathbf{p}_u\prime$ and $\mathbf{q}_i\prime$ to estimate the prediction score between the target user $u$ and the target item $i$.
\end{itemize}
The performance of COMET, COMET-original only, and COMET-interaction only is reported in Figure 4. Obviously, the combination of the original representation and the interaction vectors achieves the best performance among the three models. 
By enriching the original representations with internal interaction vectors, the latent representations of users and items are dependently learnt in COMET.  
This observation demonstrates a promising way to improve the implicit recommendation task without any additional data such as text reviews, social networks, and knowledge graphs \cite{sun2019research}.

\subsection{Sensitivity Analysis (The Answer to RQ4)}
Here we investigate the effect of the regularization parameter, the embedding size $K$, and the number of channels applied on COMET.
From Figure 6, we can observe that COMET obtains better performance with the increase in embedding size. It is reasonable since a larger embedding size is able to encode richer representations of users and items.
Besides, we can observe that COMET generally achieves better performance in the range [5e-6, 1e-5].
This indicates that it is important to choose a suitable regularization parameter to balance between overfitting and underfitting.
At last, we find that the performance of COMET based on different numbers of channels is very stable, which shows the strong expressiveness of CNN. We only conduct a sensitivity test on 3 different numbers of channels here, since the training time will increase dramatically with more channels under our setting. 
In addition, to show the convergence of our proposed method, we plot the training loss in each training epoch in Figure 7.
\begin{figure*}[t]
\centering

    \begin{minipage}[b]{0.32\textwidth}
      \centering
      \includegraphics[width=2.1 in]{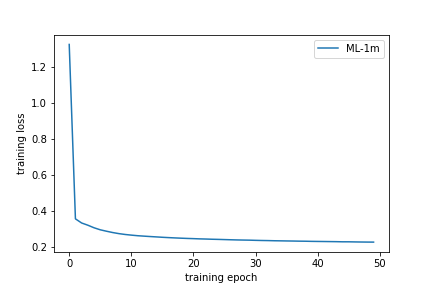}
    \end{minipage}
    \begin{minipage}[b]{0.32\textwidth}
      \centering
      \includegraphics[width=2.1 in]{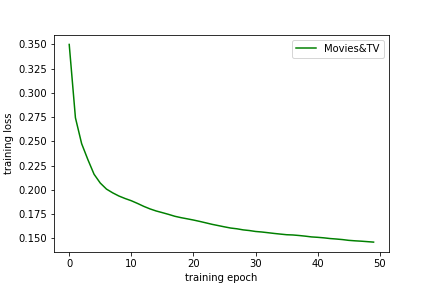}
    \end{minipage}
    \begin{minipage}[b]{0.32\textwidth}
      \centering
      \includegraphics[width=2.1 in]{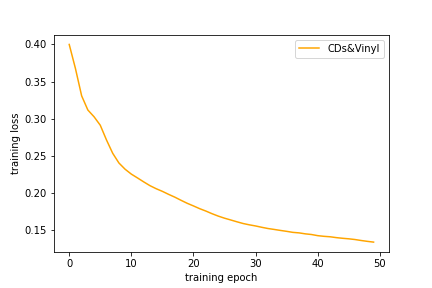}
    \end{minipage}
  \caption{Training loss of COMET on three datasets.}
\end{figure*}

\section{Conclusion}
The representation learning-based recommendation models aim to learn effective representations of users and items. In this paper, we studied how the interactions among historical interactions and embedding dimensions enrich the representations learnt by the MF.
By representing the interacted items and interacted users as two "embedding maps", COMET is able to exploit high-order interaction signals among historical interactions and embedding dimensions simultaneously. 
The advantage of enriching the representations of users and items by the learnt interaction information is also demonstrated.   
Extensive experiments and ablation studies demonstrate the efficacy of our proposed method over the existing state-of-the-art methods.
In future work, we will explore more efficient and scalable approaches to capture the interactions among historical interactions and embedding dimensions. 
In addition, we would like to apply this idea to sequential recommendation tasks and investigate the dimensional interaction effects in sequential settings.

\section{acknowledgment}
This work was supported in part by the A*STAR-NTU-SUTD AI Partnership Grant RGANS1905, and in part by the Singapore Institute of Manufacturing Technology-Nanyang Technological University (SIMTech-NTU) Joint Laboratory and Collaborative Research Programme on Complex Systems.
\bibliographystyle{ACM-Reference-Format}

\end{document}